\documentclass[aip,amsmath,amssymb,reprint]{revtex4-1}
\usepackage{color}
\usepackage{graphicx}
\usepackage{dcolumn}
\usepackage{bm}
\usepackage[mathlines]{lineno}

\usepackage[utf8]{inputenc}
\usepackage[T1]{fontenc}
\usepackage{mathptmx}
\usepackage{multirow}
\begin{document}

\preprint{AIP/123-QED}

\title{Silicon-vacancy color centers in Si- and Si,P-doped nanodiamonds: thermal susceptibilities of photo luminescence band at 740 nm}

\author{Sumin Choi}

 \affiliation{School of Mathematics and Physics, The University of Queensland, QLD 4072, Australia}
\author{Viatcheslav N. Agafonov}
\affiliation{GREMAN, UMR CNRS-7347, University F. Rabelais, 37200 Tours, France}
\author{Valery A. Davydov}
\affiliation
{L.F. Vereshchagin Institute for High Pressure Physics,  The Russian Academy of Sciences, Troitsk, Moscow 108840, Russia}
\author{L.F. Kulikova}
\affiliation
{L.F. Vereshchagin Institute for High Pressure Physics,  The Russian Academy of Sciences, Troitsk, Moscow 108840, Russia}

\author{Taras Plakhotnik}
\email{taras@physics.uq.edu.au}
\affiliation{School of Mathematics and Physics, The University of Queensland, QLD 4072, Australia}

\date{\today}

\begin{abstract}
We  have characterized  thermal  susceptibilities  of the spectral band at 740 nm of silicon-vacancy (SiV) centers in Si- and Si,P-doped nanodiamonds over  a temperature range from 295\,K to 350\,K, which is of interest for thermometry in biological systems. Si-doped crystals reveal linear dependence of the SiV zero-phonon line position,  width and relative amplitude  with susceptibilities  of 0.0126(4)\,nm/K,  0.062(2)\,nm/K and $-0.037(2)$\,K$^{-1}$, respectively.  Si,P-doped nanodiamonds show significantly  smaller (up to 35\% for the width) susceptibilities and prove  control of SiV properties  with additional chemical doping. It is argued that a significant contribution to the heating of the nanodiamonds induced by laser light   can be intrinsic due to a high concentration  and   low luminescence  quantum yield of SiV centers.  
\end{abstract}

\maketitle

Minimally invasive temperature sensing with high precision and accuracy is in demand  for  various applications including cell biology\cite{baffou2014critique}. In response, non-invasive optical nanothermometry has emerged\cite{jaque2012luminescence,brites2012thermometry,vetrone2010temperature,kucsko2013nanometre,plakhotnik_nanotech2015,toyli2013fluorescence,nguyen2018all,fan2018germanium,alkahtani2018tin}  where temperature is  measured ultra locally using fluorescent indicators.  Addition of  various chemical elements such as nitrogen\cite{toyli2013fluorescence}, silicon\cite{nguyen2018all}, germanium\cite{fan2018germanium}, and tin\cite{alkahtani2018tin}  to diamond  creates luminescent and thermo-sensitive centers.  In particular, silicon-vacancy (SiV) centers have been proposed for all-optical sensing\cite{neu2012photophysics,jantzen2016nanodiamonds,li2016nonblinking}   due to their brightness, photostability\cite{merson2013nanodiamonds}, and strong, easily detectable at room temperature zero-phonon line (ZPL). Moreover, diamond is a convenient probe delivery vehicle due to its biocompatibility\cite{zhu2012biocompatibility,zhang2013surfactant} and easiness of surface functionalization\cite{liu2004functionalization}. 

Temperature-dependent luminescence properties of SiV centers have  been studied\cite{lagomarsino2015robust,jahnke2015electron,dragounova2017determination,nguyen2018all, neu2013low}, but measured in bulk diamond or in nanodiamonds  below room temperature\cite{neu2013low}. Some of the results are also controversial. The reports on optical thermometry with SiV-centers use mostly ZPL peak position for sensing\cite{alkahtani2018tin,nguyen2018all} although multiparametric analysis using other parameters such as ZPL relative amplitude and full-width at half maximum (FWHM)  can significantly reduce the noise floor and thus improve the precision of measurements\cite{choi2019ultrasensitive}. 

In this Letter, we present temperature-dependent optical characterisation of SiV color centers in nanodiamonds at temperatures from 295 $\mathrm{K}$ to  350 $\mathrm{K}$. We investigate two types of samples: Si-doped and Si,P-doped nanodiamonds. Phosphorus (P) has attracted attention because it was considered as a candidate for n-type doping \cite{koizumi1998growth,prins1995ion}, even though some reports\cite{jones1996limitations} point on drawbacks. Here we test the idea of tuning thermal susceptibility of SiV centers, an important factor affecting precision of the temperature measurements,  by adding another impurity to Si-doped diamond.    

\begin{figure}
\centering
\includegraphics[width = 8.5cm]{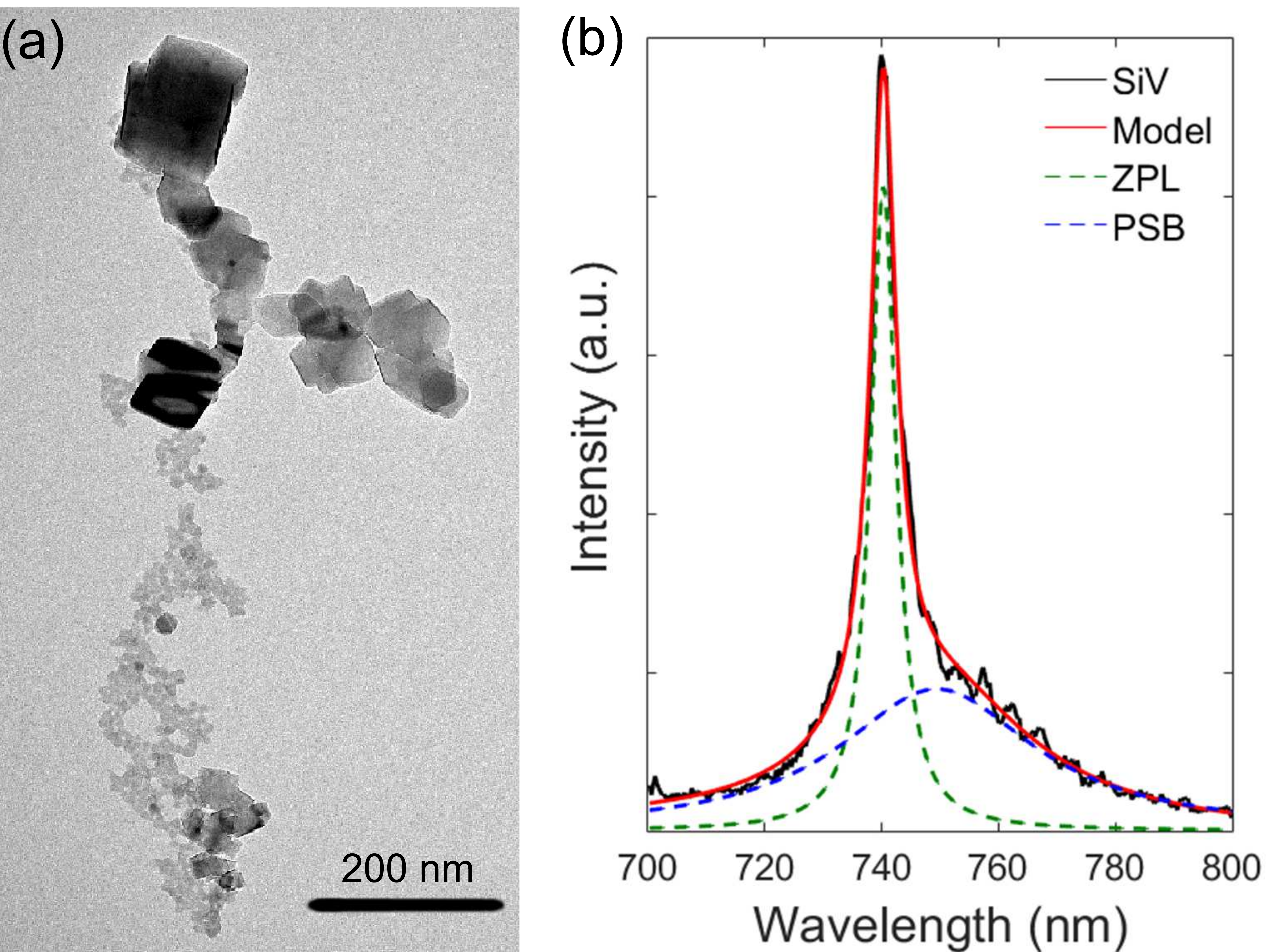}
\caption{\label{fig:figure1} (a) TEM image of nanodiamonds. (b) Photoluminescence spectrum of SiV-centers. Black dots show experimental data.  The red curve is the best fit of the function defined by Eq.\,(\ref{eq:spectrum}) to the experimental data. The green and blue dash curves are the zero-phonon line (ZPL) and the phonon side band (PSB) respectively.}
\end{figure}

Si- and Si,P-doped nanodiamonds with SiV color centers were obtained by high pressure - high temperature (HPHT) treatment of the catalyst metal-free mixtures of naphthalene - $\mathrm{C_{10}H_{8}}$ (Chemapol) with tetrakis(trimethylsilyl)sylane – $\mathrm{C_{12}H_{36}Si_{5}}$ (Stream Chemicals Co.) and naphthalene with tetrakis(trimethylsilyl)sylane and triphenylphosphine – C18H15P (Sigma-Aldrich), respectively\cite{davydov2014production}. Cold-pressed tablets of the initial homogeneous mixture (5-mm diameter and 4-mm height) were placed into a graphite container, which simultaneously served as a heater of the high-pressure toroid-type apparatus. The experimental procedure consisted of loading the high-pressure apparatus to 8.0 GPa at room temperature, heating the sample to the temperature of diamond formation (1400 $\mathrm{^{o}C}$), and short (3-10 s) isothermal exposure of the sample at this temperature. The obtained high-pressure states have been isolated by quenching to room temperature under pressure and then complete pressure release. X-ray diffraction (XRD) and Raman spectroscopy, and transmission electron microscopy were used for preliminary characterization of synthesized diamond materials.  The obtained nanodiamond products consist of different particle size distributions. A typical transmission electron microscope (TEM) image of nanodiamonds is shown in Fig.\,\ref{fig:figure1}(a). 

The nanodiamonds were dispersed in de-ionised water, treated with ultrasound  (QSonica, Q125) for 30 seconds to break down aggregates and drop-cast on a cover glass for optical characterisation.
\begin{figure}
\centering
\includegraphics[width = 9cm]{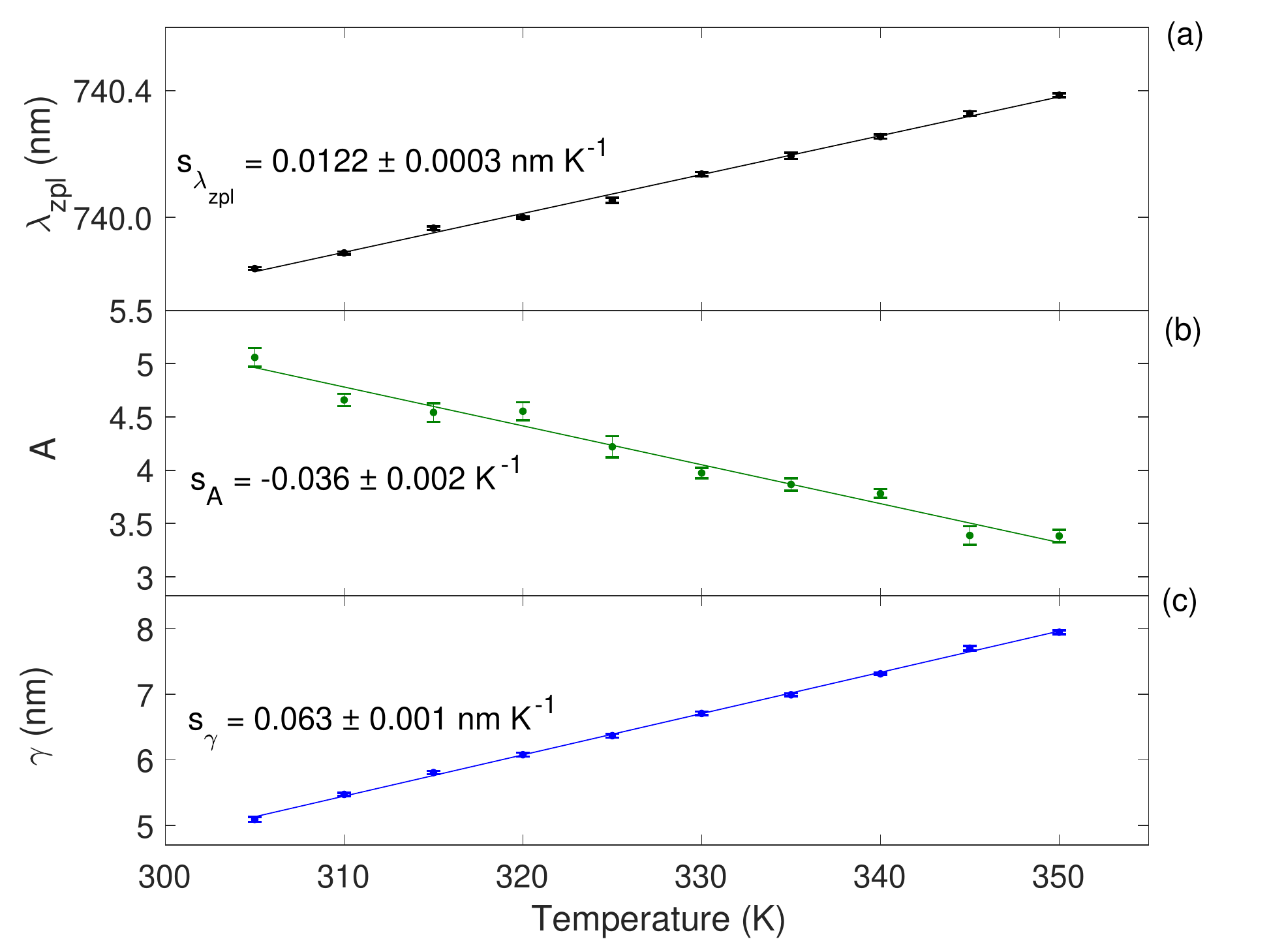}
\caption{\label{fig:SiV_temp}  Effects of temperature on optical properties of SiV centers in Si-doped nanodiamonds. (a) ZPL position $\lambda_\mathrm{zpl}$ , (b) Relative amplitude $A$, and (c) ZPL FWHM $s_\gamma $. Thermal susceptibilities are displayed in the corresponding panels. }
\end{figure}
Optical properties of the Si- and Si,P-doped nanodiamonds were studied with a home-built microscope. Luminescence  of the SiV centers was excited by a 532-nm  laser (Cohrernt, Verdi-V5) focused onto a single crystal (an object with a diffraction limited size of the optical image) through an air objective (Nikon, NA = 0.55). The temperature was set using a Peltier element heating/cooling stage  (TC-720 Thermoelectric Temperature Controller). The emission collected by the same objective  and filtered to reject the excitation light was  directed to a spectrometer (Princeton Instruments Acton 2300i). To minimize effects of spatial displacement of the sample (particularly after every change of the stage temperature) on the accuracy of spectral measurements, we employed a grating with 1200 l/mm ruling which provided a large dispersion (about 2.05 nm/mm). 

Photo luminescence spectra revealed bands with ZPLs at 740-nm wavelength\cite{SiV_ZPL} and red-shifted PSBs, Fig.\,\ref{fig:figure1}(b). The shape of the luminescence band $\Phi(\lambda)$ is modeled by a sum of two Lorentzians and a linear background term $B(\lambda)$ 
\begin{equation}\label{eq:spectrum}
\Phi(\lambda)=\frac{ \gamma^2A_\mathrm{zpl}}{4(\lambda-\lambda_\mathrm{zpl})^2+\gamma^2}+\frac{\Gamma^2A_\mathrm{psb}}{4(\lambda-\lambda_\mathrm{psb})^2+\Gamma^2}+B(\lambda)
\end{equation} 
Separation of the spectrum into the ZPL and the PSB is ambiguous and the true shape of the two parts is more complicated but Eq.\,(\ref{eq:spectrum}) satisfactorily fits the data as can be seen in Fig.\,\ref{fig:figure1}(b).  The following parameters -- ZPL width $\gamma$, position $\lambda_\mathrm{zpl}$, relative amplitude $A \equiv A_\mathrm{zpl}/A_\mathrm{psb}$,  PSB width $\Gamma$, and splitting between ZPL and PSB maxima $\lambda_\mathrm{psb}-\lambda_\mathrm{zpl}$  were extracted from  spectra measured at different temperatures by fitting Eq.\,(\ref{eq:spectrum}) to the experimental data using a least-squares method. Thermal susceptibility of $\gamma$ (similar for other parameters) is defined as follows  $s_\gamma \equiv \partial \gamma/\partial T$. 

The temperature of the Peltier stage has been variable from 293\,K to 318\,K with an increment of 5\,K but the actual temperature of the nanodiamonds  was higher at each setting because of heating caused by the laser light.  To take this into account, we measured spectra at two laser powers, 20 mW and 50 mW and the value of $s_\gamma$ has been determined at each power. It has been found that  $s_\gamma$ does not depend on laser power or stage temperature  within the investigated range and the accuracy of measurements.   The difference  between $\gamma$ at 50\,mW and 20\,mW and the value of $s_\gamma$ have been used to find $\Delta T_{30}$ for each crystal, the increment of temperature caused by  30-mW increase of the laser power. To obtain the actual crystal temperature at 20 mW and 50 mW,  $2/3\times \Delta T_{30} $ and $5/3\times \Delta T_{30} $ were added to the stage temperature  respectively.  As an example, data with corrected temperature are shown in Fig.\,\ref{fig:SiV_temp}. First four data points are measured at 20 mW (305 K to 320 K), and the remaining six  are obtained at 50 mW. Similar procedure has been applied to Si,P-doped diamond.  We have summarized all the observations  for all measured crystals in Tab.~\ref{tab:table1} and Tab.~\ref{tab:table2}.  These tables also include $R$, the total detected photon count rate at 50 mW. 

On average, the temperature increase was about 13\,K and 3\,K for Si-doped and Si,P-doped nanodiamonds. The origin of heating is not clear but  the heat generated directly by SiV centers needs a due consideration. The luminescence quantum yield of SiV centers is about 5\% in a CVD diamond film\cite{SiV_Qyield} but it can be  several times smaller in nanocrystals\cite{plakh2018}. The power of heat released by SiV centers inside a diamond crystal is $W=Rhc/(\lambda QD)\approx 75$\,nW, where quantum yield  $Q\approx 0.01$, photon detection efficiency\cite{NC_quantative} $D\approx0.02$, and detected photon count rate $R \approx40$\,MHz.  The corresponding steady-state increase of temperature  $\Delta T = W/(4\pi \kappa r)$ is about 4\,K under the assumptions that the heat conductivity coefficient $\kappa$ is that of air and the radius of the crystals $r\approx50$\,nm. This $\Delta T$ is an order of magnitude smaller than the value observed in the experiment but the effective heat conduction from the particle to the air is much smaller than obtained using the bulk value of $\kappa$ when the size of the crystal in comparable  to the mean free path of air molecules\cite{heat_small} (about 70 nm). Thermal resistance between  the substrate and the crystal may also be significant and dependent on the shape and orientation of the nanocrystal.  Thus, it is plausible that not radiative energy release by photo excited SiV centers is responsible for the heating. 
\begin{figure}
\centering
\includegraphics[width = 9cm]{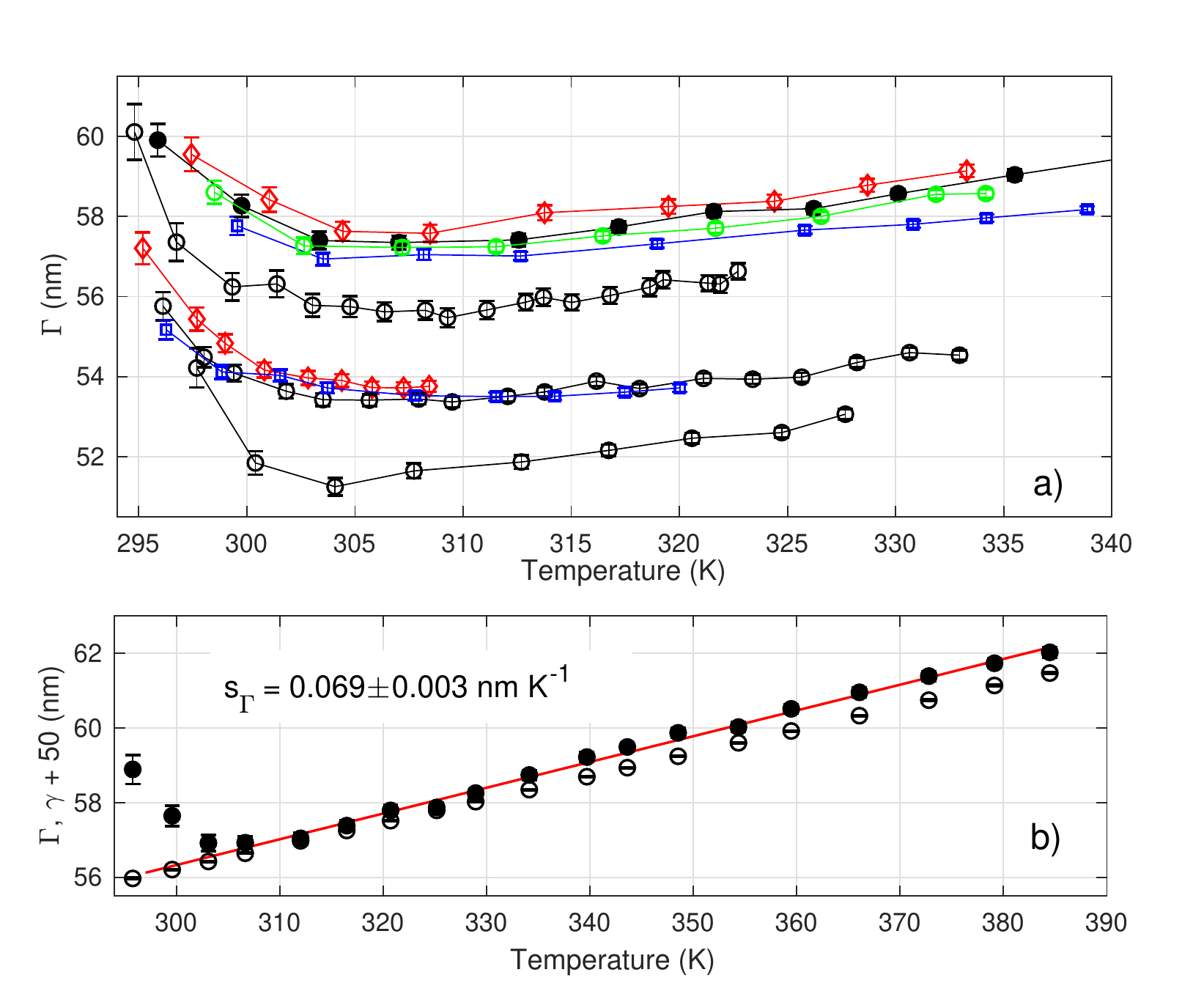}
\caption{\label{fig:PSB_FWHM} FWHM of the PSB in Si-doped diamond. a) Obtained at different laser powers and the corresponding temperature has been derived from the change of $\gamma$. The  lines between points are for guidance only. b) $\gamma +50\text{\,nm}$ (open circles) and $\Gamma$ in a wider range for  one of the crystals. Solid line is a linear fit to data above 305\,K. }
\end{figure}

Temperature dependence of two remaining parameters of SiV spectra, the PSB peak position and width were relatively harder to measure accurately. This is because they turned out to be much more affected by small changes in the position of the crystal  due to thermal expansion of the stage.  To overcome this difficulty,  we have used  laser power to regulate crystal temperature and kept  the stage temperature constant.  The results are plotted in Fig.\,\ref{fig:PSB_FWHM}  for several nanocrystals showing a common feature, a minimum at approximately 305\,K and a nearly linear increase above where the value of $s_\Gamma \approx 0.069$\,nm/K  is similar to the value of $s_\gamma$. The origin of the initial decrease is unclear but positions of the minima in Fig.\,\ref{fig:PSB_FWHM} will not be aligned horizontally if power is used  instead of temperature as a Cartesian coordinate in Fig.\,\ref{fig:PSB_FWHM}.  

Susceptibilities $s_{\lambda_\mathrm{zpl}}$ and $s_\gamma$ in Si-doped diamond are consistent within the standard deviation with previous measurements in bulk\cite{nguyen2018all, lagomarsino2015robust}.  Si,P- and Si-doped nanodiamonds show similar $s_{\lambda_{zpl}}$ while $s_A$ and especially $s_\gamma$ are significantly smaller (up to 35\%) in Si,P-doped nanodiamonds.  These results prove that additional chemical doping may significantly affect the thermal properties of SiV centers and thus a new direction for research opens up. The sample standard deviations   in Si,P-doped nanodiamonds are approximately 3 times larger than in Si-doped samples. A smaller number of SiV centers per crystal in  Si,P-doped nanodiamonds as suggested by smaller photon rates $R$ can explain the difference. A smaller size of the Si,P-doped nanodiamonds and/or  smaller Si concentration (because P and Si atoms are competing for analogous sites in the crystal lattice) are the factors reducing the number of SiV centers. 

In summary, we have characterized effects of temperature on luminescent properties of  SiV centers in Si- and Si,P-doped nanodiamond samples. In Si-doped crystals and in the temperature range  from 305\,K to 350\,K, susceptibilities $s_{\lambda_\mathrm{zpl}}$, $s_A$, and $s_\gamma$ are 0.0126(4)\,nm/K, $-0.037(2)$\,K$^{-1}$ and 0.062(2)\,nm/K, respectively. Simultaneous introduction of Si and P in nanodiamonds affects luminescence spectra of SiV centers and their dependence on temperature.  In particular, $s_{\lambda_{zpl}}=0.0140(8)$\,nm/K, $s_A = -0.031(1)$\,K$^{-1}$ and $s_\gamma =  0.046(2) $\,nm/K in Si,P-doped crystals. Decrease of $s_\gamma$ is quite significant. Thus  temperature susceptibility of SiV centers can be modified by co-doping and further research may identify elements significantly enhancing the temperature response. High concentration of SiV can result in intrinsic heating which should be carefully considered.  This work provides  comprehensive characterization of luminescent properties of SiV centers in nanodiamonds which are important for optical thermometry especially in the range relevant for biological applications. \\

The financial support is provided by Human Frontier Science Program, RGP0047/2018.  V.\,D.  thanks Russian  Foundation for Basic Research (Grant 18-03-00936).
 
\begin{table*}
\caption{\label{tab:table1} Parameters  characterizing the luminescence band of Si-doped nanodiamonds at $\approx 325$\,K  and  thermal susceptibilities. }
\begin{ruledtabular}
\begin{tabular}{c|cccccccccc}
&\multicolumn{10}{c}{$\textnormal{Si-doped nanodiamond}$}\\
\hline
\multicolumn{1}{c|}{\multirow{2}{*}{No.}} & $\lambda_\mathrm{zpl}$ & $s_{\lambda_\mathrm{zpl}}$  & $\gamma$ & $s_{\gamma}$ & $A$ &  $s_A$ & $\Gamma$ & $\lambda_\mathrm{zpl}-\lambda_\mathrm{psb}$ & $\Delta T_{30} $ & $R$ \\  
\multicolumn{1}{c|}{} & nm  & $\mathrm{nm/K}$ & nm  & $\mathrm{nm/K}$ &{} & $\mathrm{K}^{-1}$  & nm & nm &K& MHz \\ 

\hline
1& 740.179(6) & 0.0114(5)& 7.13(2) & 0.059(2)  & 4.47(3) & -0.031(3)  & 58.6(8) &  10.0(1) & 23.3& 46.8  \\
2& 740.109(5) & 0.0114(8)& 6.47(1) & 0.059(1)  & 4.00(8) & -0.042(6)  & 62(1) & 9.4(2)  & 23.1& 34.0 \\
3 & 740.347(1) & 0.012(1) & 6.44(2) & 0.064(3) & 4.03(2) & -0.033(3)  & 66.5(6) & 9.4(2) & 21.9& 31.8  \\
4& 740.053(8) & 0.0131(6) & 6.34(3) & 0.064(2) & 4.22(9) & -0.035(4)  & 59(1) & 12.8(1) & 20.0& 41.0 \\
5 & 740.238(6) & 0.0130(4)& 6.28(2) & 0.066(1) & 4.34(3) & -0.042(3)  & 67.2(5) & 9.9(2)  & 19.1& 40.1 \\
6 & 740.467(5) & 0.014(1) & 6.82(2) & 0.053(5)& 3.99(2) & -0.037(3)  & 62.3(4) & 11.3(1)   & 19.9& 100.9 \\
7& 740.256(6) & 0.012(2) & 6.26(1) & 0.066(2) & 4.57(3) & -0.041(4)  & 64.0(4) & 13.2(1)  & 17.2& 28.2 \\
8 & 740.194(7) & 0.0114(4)& 6.04(1) & 0.062(3) & 4.55(3) & -0.032(3)  & 64.8(5) & 13.5(1)  & 21.2& 38.9\\
9 & 739.872(6) & 0.015(1)& 6.36(1) & 0.067(4) & 4.56(8) & -0.042(5)  & 49.3(2) & 11.6(1)  & 18.1& 37.0\\
10& 740.247(3) & 0.0120(6) & 6.70(1) & 0.057(4) & 4.09(6) & -0.035(3)  & 49.9(4) & 10.4(1)  & 20.9& 16.8 \\
\hline
Mean  & 740.20 & 0.0126 & 6.49 & 0.062  & 4.28  & -0.037 & 60.3 & 11.5  & 20.5&  41.5 \\  
std  & 0.16 & 0.0013 & 0.32 & 0.0047 & 0.25 & 0.0044 & 6.3 & 1.5  & 2.0& 22.4\\
\end{tabular}
\end{ruledtabular}

\end{table*}

\begin{table*}
\caption{\label{tab:table2} Parameters  characterizing the luminescence band of Si,P-doped nanodiamonds at $\approx 298$\,K and thermal susceptibilities. } 
\begin{ruledtabular}
\begin{tabular}{c|cccccccccc}
&\multicolumn{10}{c}{$\textnormal{Si,P-doped nanodiamond}$}\\
\hline
\multicolumn{1}{c|}{\multirow{2}{*}{No.}} & $\lambda_\mathrm{zpl}$ & $s_{\lambda_\mathrm{zpl}}$  & $\gamma$ & $s_{\gamma}$ & $A$ &  $s_A$ & $\Gamma$ & $\lambda_\mathrm{zpl}-\lambda_\mathrm{psb}$ & $\Delta T_{30}$  &   $R$ \\  
\multicolumn{1}{c|}{} & nm  & $\mathrm{nm/K}$ & nm  & $\mathrm{nm/K}$ &{} & $\mathrm{K}^{-1}$  & nm & nm &K& MHz \\ 
\hline
1 & 739.872(3) & 0.015(1) & 6.65(2) & 0.062(4)& 4.54(4) & -0.028(5) & 54.5(8) & 11.4(2) & 6.1 & 0.6 \\
2 & 739.993(4) & 0.054(7)& 6.38(2) & 0.042(1) & 4.56(5) & -0.033(4) & 33.0(4) & 10.6(1)  & 6.9& 1.8\\
3& 739.848(3) & 0.010(1)& 4.94(1) & 0.037(2)  & 6.36(7) & -0.035(4) & 56(1) & 11.6(3)  & 2.5& 5.3 \\
4 & 739.598(3) & 0.021(2)&5.34(1) & 0.061(1) & 5.38(5) &  -0.027(4) & 57.3(6) & 13.1(2)  & 8.5& 1.8 \\
5 & 739.771(4) & 0.0145(6) & 4.83(1) & 0.043(2) & 6.31(5) & -0.031(4) & 53.9(5) & 10.7(1)  & 4.6& 5.5 \\
6 & 740.476(7) & 0.014(2) & 6.93(2) & 0.059(3) & 3.98(3) & -0.022(3) & 43.5(4) & 15.0(1)  & 7.7& 5.9 \\
7& 739.639(6) & 0.010(1) & 6.15(1) & 0.033(2) & 5.53(4) & -0.030(3) & 55.7(5) & 13.9(2)  & 11.2& 3.9 \\
8 & 739.836(6) & 0.014(2) & 5.98(2) & 0.036(4) & 4.03(3) & -0.034(4) & 63(1) & 16.4(3)  & - & 0.4\\
9 & 740.251(7) & 0.0139(8) &6.03(1) & 0.048(2) & 4.81(3) & -0.022(3) & 49.2(5) & 11.2(2) & 9.7& 1.8 \\
10& 740.358(7) & 0.0138(6) & 6.25(1) & 0.045(3)  & 5.55(7) & -0.029(2) & 54.2(9) & 14.2(1) & 7.8 & 0.4 \\
11 & 740.238(7) & 0.014(1) & 7.53(1) & 0.049(3) & 4.24(4) & -0.027(3) & 47.3(5) & 10.2(2) & 8.3 & 1.3 \\
12 & 740.478(8) & 0.013(2) & 6.69(2) & 0.050(2) & 3.92(3) & -0.030(3) & 47.3(5) & 17.2(1)  & 7.7& 5.3 \\
13 & 739.579(3) & 0.0130(4) & 4.75(1) & 0.045(1) & 6.45(5) & -0.031(3) & 56.6(7) & 13.8(2)  & 7.6& 1.8 \\
14 & 739.872(2) & 0.0132(6) & 5.09(2) & 0.042(1)  & 5.10(5) & -0.032(4) & 53.6(7) & 12.7(2)  & 8.0& 3.8 \\
\hline
Mean  & 739.0 & 0.014 & 5.96 & 0.046  & 5.05 & -0.031 & 51.9 & 13.0  & 7.4 & 2.8\\  
std  & 0.32 & 0.0030  & 0.86 & 0.0091 & 0.90 & 0.0037 & 7.4 & 2.2 & 2.0 & 2.0\\

\end{tabular}
\end{ruledtabular}
\end{table*}
\nocite{*}
%

\end{document}